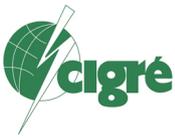

# Load Curtailment Estimation in Response to Extreme Events

R. ESKANDARPOUR, A. KHODAEI   A. ARAB
University of Denver   Protiviti
USA   USA

## SUMMARY

A machine learning model is proposed in this paper to help estimate potential nodal load curtailment in response to an extreme event. This is performed through identifying which grid components will fail as a result of an extreme event, and consequently, which parts of the power system will encounter a supply interruption. The proposed model to predict component outages is based on a Support Vector Machine (SVM) model. This model considers the category and the path of historical hurricanes, as the selected extreme event in this paper, and accordingly trains the SVM. Once trained, the model is capable of classifying the grid components into two categories of outage and operational in response to imminent hurricanes. The obtained component outages are then integrated into a load curtailment minimization model to estimate the nodal load curtailments. The merits and the effectiveness of the proposed models are demonstrated using the standard IEEE 30-bus system based on various hurricane path/intensity scenarios.

Rozhin.Eskandarpour@du.edu
This work has been supported in part by the U.S. National Science Foundation under Grant CMMI-1434771

# 1. INTRODUCTION

Resilience is defined as the rate and speed of a system in returning to normal conditions after an external shock [1]. Among all types of extreme events, hurricanes are notably recognized as the most recurring in the United States, mostly occurred by the Atlantic Ocean and Gulf of Mexico [3]. An accurate forecast of the component outage and load curtailment in response to this extreme event is an essential task in pre-event and post-event planning/recovery of power systems. Improving resilience in power systems is extensively discussed in the literature and there are several works on system modelling, resource allocation and optimal scheduling for enhancing grid resilience. In [3] a proactive resource allocation method is proposed to repair and recover power grid components after an extreme event. A proactive recovery framework of the components and a deterministic recovery model are proposed in [4] and [5] to manage the available resources before and after an extreme event. In [6], a restoration model is proposed by considering AC power flow and identifying the macroeconomic concept of the value of lost load (VOLL) in an optimal scheduling problem to find the minimum economic loss in case of load interruptions.

In various research efforts, a proper formulation of a problem and its closed solution cannot be easily found. Machine learning approaches are capable to learn and forecast from historical data. The historical data can be categorized by a supervised learning algorithm (classification), unsupervised learning (clustering), or different prediction methods (mostly regression modelling) [7]. In the power and energy research, various machine learning methods are utilized to solve a variety of problems [8]. The application of machine learning in power grids include risk analysis using regression models and artificial neural networks (ANNs) [9], distribution fault detection using ANNs and Support Vector Machine (SVM) [10], to name a few. There are also few machine learning works to estimate component outages in response to extreme events. In [11], a logistic regression model is applied considering the wind speed and the distance of the each component from the center of the hurricane as two major features to find the state of the each component after an extreme event.

In this paper, a Support Vector Machine method is used to determine the state of each component in response to an imminent hurricane. The predictions are then integrated into a minimum load curtailment model to calculate potential nodal load curtailments, which can be useful for grid operators to identify critical and prone-to-curtailment areas in the system and to further allocate resources and repair crews before/after the extreme event. The rest of the paper is organized as follows: Section 2 presents the problem statement and proposes two models for outage prediction and load curtailment estimation. Section 3 presents a case study of using the proposed Support Vector Machine model and estimating the nodal load curtailment using IEEE 30-bus test system. Finally, Section 4 concludes the paper.

# 2. COMPONENT OUTAGE PREDICTION AND LOAD CURTAILMENT ESTIMATION

Fig. 1 indicates the outline of the proposed model. The problem is solved in three consecutive stages. In Stage (a), the category and the path of an upcoming hurricane are predicted. The category and path are used to identify the intensity of the hurricane and the potentially impacted regions, respectively. These data are obtained from weather forecasting agencies. In Stage (b), the speed of the hurricane, and the distance of each power grid component from the center of the hurricane, which are represented by parameters $x_1$ and $x_2$ respectively, are used to predict the state of a component. A SVM method is used in this stage to classify the components into two states of damaged (on outage) and operational (in service). The SVM model is trained on historical data. Stage (c) solves a minimum load curtailment problem considering the predicted state of each component to estimate the potential nodal load curtailments.

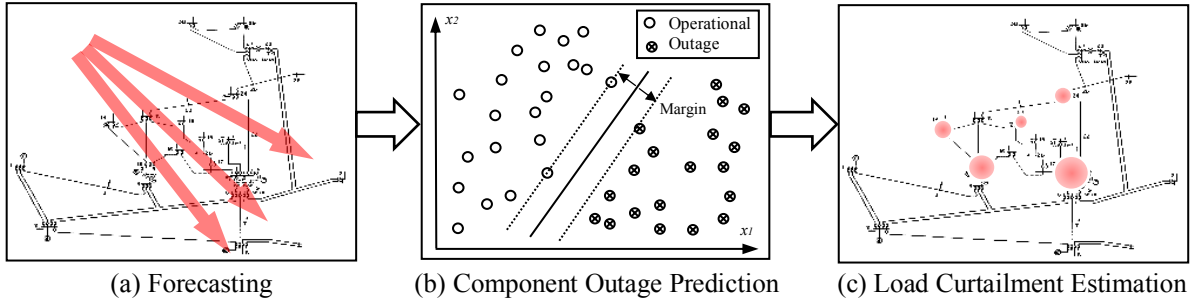

(a) Forecasting  (b) Component Outage Prediction  (c) Load Curtailment Estimation

Fig. 1 Proposed model

## 2.1. Component Outage Prediction

Figure 1.b, illustrates a schematic of the damaged (cross) and operational (circle) components based on distance and wind speed (intensity of the hurricane), separated by a decision boundary. First, a machine learning method is used and trained to determine the decision boundary; subsequently, power grid component outages in response to upcoming hurricanes can be effectively predicted. A Support Vector Machine (SVM) method [12] is applied to classify the component states using a decision boundary into two classes (operational and outage) based on the wind speed ($x_1$) and the distance of each component from the center of the hurricane ($x_2$), Figure 1.b illustrates the obtained support vectors and optimal hyperplane separating two classes.

An SVM is a binary classifier that separates training examples of one class from the other by defining a proper hyperplane. The best hyperplane is defined as the hyperplane with the widest gap, obtained by a quadratic programming problem (1):

$$\min \ \frac{1}{2}\|w\|^2 + c\sum_{i=1}^{m}\varepsilon_i \quad (1)$$

s.t.
$$y^{(i)}\left(w^t x^{(i)} + b\right) \geq 1, -\varepsilon_i, \quad i=1,\ldots,m$$
$$\varepsilon_i \geq 0, \quad i=1,\ldots,m$$

where $y^{(i)}$ represents the class labels of the training examples (i.e., -1 and +1 in case of a binary classification problem), $x^{(i)}$ is the feature vector of each training example, $w$ is the normal vector to the hyperplane separating training examples, and $|b|/\|w\|$ is the perpendicular distance from the hyperplane to the origin. This quadratic programming problem can be solved by a Lagrange duality.

The quadratic programming problem (1) is developed based on the assumption that classes are linearly separable. In a case that the training data cannot be separated by a linear hyperplane (which is common), SVM can use a soft margin. The soft margin classification is solved by introducing a penalty parameter $c$ and a regularization (often L1 or L2). $\varepsilon_i$ is the regularization weight of the samples in the margin (support vectors). In other words, if an example has functional margin 1-$\varepsilon_i$ (with $\varepsilon_i$ >0), the objective function is penalized by $c\varepsilon_i$.

Another approach of applying SVM to nonlinear data is the kernel method [12]. The idea of a kernel method (or as sometimes called kernel trick) is to map the input feature vector into a higher-dimension space where the classes are linearly separable. Kernel trick simply states that inner product of $x_1$ and $x_2$ in the input space can be replaced by a certain function $K(x_1,x_2)$. For example, a polynomial kernel of degree $d$ can be defined as: $K(x_1,x_2) = (x_1,x_2)^d$. Finding a proper value of penalty parameter $c$ and the best kernel depends on the shape of classes, which are often unknown. Therefore, $c$ and the kernel function are often found via a search method to minimize the error on a test set.

## 2.2. Load Curtailment Estimation

The objective of the minimum load curtailment problem is defined as the value-weighted cost of load curtailment in the system as in (2):

$$\min \sum_t \sum_s \sum_b VOLL_b \times LC_{bts} \qquad (2)$$

where $LC_{bts}$ is the amount of load curtailment at bus $b$ at time $t$ during contingency scenarios $s$. The Value of Lost Load (VOLL) illustrates the average cost that each customer is willing to pay in order to avoid any load interruptions [13]. Assuming $UX_{its}$ is the outage state of unit $i$ at time $t$ in scenario $s$ (considering operating state equals to 1 and outage state equals to 0) and $UY_{lts}$ is the outage state of line $l$ at time $t$ in scenario $s$ (considering operating state equals to 1 and outage state equals to 0), the proposed objective is subject to the following operational constraints:

$$\sum_{i \in B_b} P_{its} + \sum_{l \in B_b} PL_{lts} + LC_{bts} = D_{bt} \qquad \forall b, \forall s, \forall t \qquad (3)$$

$$P_i^{\min} I_{it} UX_{its} \leq P_{its} \leq P_i^{\max} I_{it} UX_{its} \qquad \forall i, \forall s, \forall t \qquad (4)$$

$$|P_{it0} - P_{its}| \in \Delta_i \qquad \forall i, \forall s, \forall t \qquad (5)$$

$$PL_l^{\max} UY_{lts} \leq PL_{lts} \leq PL_l^{\max} UY_{lts} \qquad \forall l, \forall s, \forall t \qquad (6)$$

$$\left| PL_{lts} - \frac{\sum_b a_{lb} \theta_{bts}}{x_l} \right| \leq M(1 - UY_{lts}) \qquad \forall l, \forall s, \forall t \qquad (7)$$

where $b$, $i$, and $l$ are the indices of buses, generation units, and lines, respectively. B is set of components connected to bus $b$, $s$ is index for scenarios, and $t$ is index for time. $P_i^{\max}$ and $P_i^{\min}$ represent the maximum and minimum generation capacity of unit $i$, respectively, $PL_{lts}$ is the real power flow of line $l$ at time $t$ in scenario $s$, $\theta_{bts}$ is the phase angle of bus $b$ at time t in scenario $s$, and M is a large positive constant. The value of $a_{lb}$ is the element of line $l$ and bus $b$ at line-bus incidence matrix, and $D_{bt}$ is the Load at bus $b$ at time $t$.

The total injected power to each bus from generation units and line flows is equal to the nodal load which can be ensured by load balance equation (3). Load curtailment variable ($LC_{bts}$) ensures a feasible solution in case of component outages when there is not sufficient generation and/or transmission capacity to supply loads. Generation unit output power is limited by its capacity limit and will be set to zero depending on its commitment and outage states (4). The change in unit generation is further limited by the maximum permissible limit between normal and contingency scenarios (5). Transmission line capacity and power flow constraints are modeled by (6) and (7), respectively, in which the outage state is included to model the line outages in contingency scenarios effectively.

## 3. CASE STUDY

As historical data for the past hurricanes at component level are limited, a set of synthetic data is generated to train the SVM model. The data includes 300 samples in outage state and 300 samples in the operational state. To define the synthetic data, Saffir-Simpson Hurricane Scale [14] is used to generate wind speed features of the synthetic data. These generated scenarios will be considered as a pre-process to the proposed machine learning method, hence generating relevant outage scenarios. A subset of data (80%) is randomly selected and used for training the SVM model, and the rest (20%) is used to validate the trained model. The output of this model, i.e., the outage state of various power grid components, can be considered as the input to the proper grid response, recovery, and scheduling models [3], [4].

In previous work a linear SVM is applied to classify the data into two classes. Although the result was accurate and reliable, the linear model is not sufficient to categorize complicated and intertwined classifications. In this work, in order to find the best kernel and its penalty parameters, different kernels (linear, polynomial quadratic, and Gaussian) with different ranges of penalty parameter ($c = 0.01, 0.1, 1, 10$) are examined. Table 1 shows the accuracy of SVM with various penalty parameters and kernels. Among the trained models, polynomial kernel SVM with $c=1$ offers the best overall classification accuracy on the validation set. The margin size of the SVM with polynomial kernel is 0.1131, and the average $\varepsilon$ (regularization weight) is 0.4558.

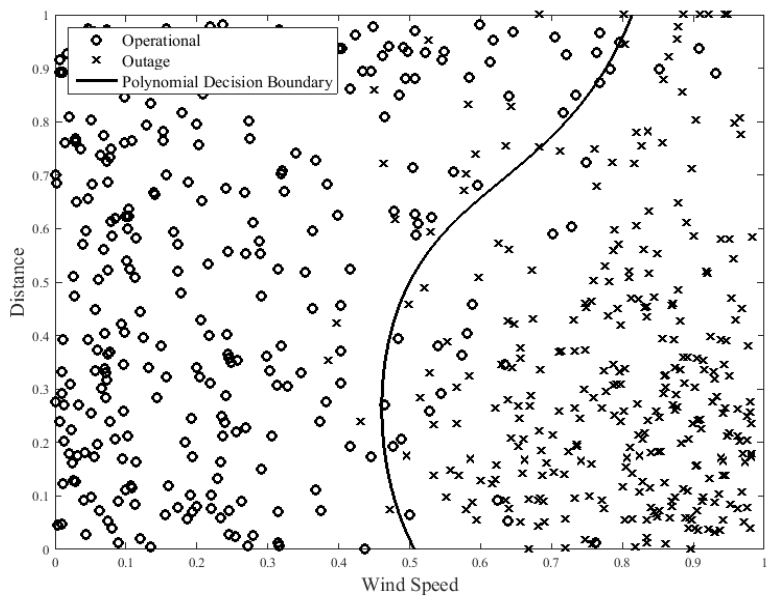

Fig. 2. Decision boundary of the polynomial kernel with penalty parameter $c=1$

Figure 2 shows the decision boundary of the polynomial kernel with penalty parameter $c=1$, separating outage from operational components based on wind speed and distance from the center of the hurricane. As shown, the data are not linearly separable, and a nonlinear kernel is necessary to better classify components. Table 2 shows the confusion matrix of this classification. As shown, the proposed method can classify the components into outage and operational with high accuracy.

Table 1. Accuracy (%) of SVM with various penalty-parameters and kernels

| Kernel | $c=0.1$ | $c=1$ | $c=10$ |
|---|---|---|---|
| Linear | 91.0 | 91.4 | 91.2 |
| Quadratic | 91.3 | 91.2 | 91.2 |
| Polynomial | 92.3 | **92.8** | 92.7 |
| Gaussian | 91.3 | 91.2 | 91.8 |

Table 2. Confusion Matrix of classifying system components

| | | Predicted | |
|---|---|---|---|
| | | Normal | Outage |
| Actual | Normal | 91.7% | 8.3% |
| | Outage | 6.0% | 94.0% |

The proposed minimum load curtailment model is applied to the standard IEEE 30-bus test system. A hurricane passes through three hypothetical paths with different intensities. Particularly, based on the available hurricane data and the estimated distance to the center of the hurricane, the state of each component in the system is found using the trained SVM model. This study will identify how much load curtailment will occur in response to an imminent hurricane. Table 3 shows the load curtailment of each contingency scenario based on the identified outages.

As shown, buses 3 and 18 are the most sensitive buses, since in both scenarios two and three there are outages in these two buses. In addition, buses 18, 19, and 20 are the most critical buses as there is approximately 96% load curtailment in these buses. The predicted outage and estimated load curtailment can be useful for utilities to allocate necessary resources and repair crews before/after the hurricane.

Table 3. Load Curtailment of Bus Outages along three Hurricane Paths

| Bus number | Total Load (MWh) | LC Scenario 1 (MWh) | LC Scenario 2 (MWh) | LC Scenario 3 (MWh) |
|---|---|---|---|---|
| 2 | 423.08 | 0 | 0 | 4.91 |
| 3 | 46.79 | 44.95 | 0 | 1.62 |
| 15 | 159.87 | 0 | 0 | 0.37 |
| 18 | 62.39 | 0 | 59.94 | 2.10 |
| 19 | 185.22 | 0 | 177.95 | 0 |
| 20 | 42.89 | 0 | 41.21 | 0 |
| 23 | 62.39 | 0 | 0 | 9.92 |
| 24 | 169.62 | 0 | 0 | 162.97 |
| 29 | 46.79 | 0 | 0 | 0.31 |

## 4. CONCLUSION

In this paper, a support vector machine was proposed to determine the probable outage of power grid components in response to an imminent hurricane. The proposed model can derive a decision boundary between operational and outage components directly from its training data. Experimental results showed the effectiveness of SVM in separating operational and outage classes from each other based on the wind speed and the distance from the center of the hurricane. A minimum load curtailment problem was subsequently proposed and formulated to estimate the amount of load curtailment given the predicted outages from the SVM model. By using the proposed model under different hurricane scenarios, the critical buses in the system more prone to experiencing an outage can be identified. Considering the importance of accurate power grid outage prediction in many applications such as operation, response, and recovery, the proposed model can be useful to power grid operators in efficiently improving grid resilience.